\documentstyle[epsf,preprint,aps]{revtex}
\draft
\begin{document}
\title{Cosmic Microwave Anisotropies from Topological Defects
in an Open Universe}
\author{Ue-Li Pen and David N. Spergel}
\address{ Princeton University Observatory}

\maketitle
\begin{abstract}

We present a general formalism for computing Cosmic Background
Radiation (CBR) and density fluctuations in open models with stiff
sources.  We decompose both the metric fluctuations and the
fluctuations in the stress-energy tensor into scalar, vector and
tensor modes.  We find analytic Green's functions for the linearized
Einstein equations in the presence of stiff sources and use this
formalism to estimate the amplitude and harmonic spectrum of microwave
background fluctuations produced by topological defects in an open
universe.  Unlike inflationary models that predict a flat universe and
a spectrum of CBR fluctuations that is enhanced at large angular
scales, defect models predict that CBR fluctuations are suppressed on
angular scales larger than that subtended by the curvature scale.  In
an $\Omega = 0.2 - 0.4$ universe, these models, when normalized to the
amplitude of CBR fluctuations observed by COBE, require a moderate
bias factor, $2-3$, to be compatible with the observed fluctuations in
galaxy counts.  In these models, accurate predictions can be made
which are testable through CBR experiments in the near future.  A CBR
measurement of $\Omega$ would then be possible, up to the limit
imposed by cosmic variance.  We discuss some of the philosophical
implications of an open model and propose a solution to the flatness
problem.

\end{abstract}
\section{Introduction}

In recent years, most theoretical work in cosmology has
assumed that the universe is flat and matter-dominated.  This
assumption is nearly inevitable in inflationary scenarios that predict that
$\Omega$ should be close to unity.  However, defect models make no
such predictions for the density of the universe.

There is a host of astronomical evidence that suggests that the
universe may be open.  White et al.\cite{whi93} argues that X-ray
observations of clusters imply that at least 20\% of their mass is
baryonic.  When combined with standard hot big-bang estimates of the
baryon density\cite{wal92}, $\Omega_b h^2 \simeq 0.015$, this implies
that the total density in non-relativistic matter is much less than
unity.  Observations of galaxy random velocities\cite{fis94a} find
that $\sigma_8 \simeq 300$ km/s, a factor of three below the
predictions of COBE normalized flat scale-invariant cosmologies.  The
matter density inferred from comparisons between the galaxy
correlation function in redshift and real space is also much smaller
than unity and is compatible with $\Omega \sim 0.3$\cite{fis94b}.

Over the past decade, observational cosmologists have devoted much
effort to measuring the spectrum of density fluctuations and have
found significant evidence for more large-scale structure than
predicted in flat universe models.  As the predicted spectrum of
density fluctuations is peaked on the physical scale corresponding to
the horizon size at matter-radiation equality, $32/\Omega h^2$ Mpc,
models with $\Omega =1$ predict less large scale power than low
$\Omega$ models.  For example, the standard CDM model underpredicts
the ratio of galaxy fluctuations observed on the 30/h Mpc scale in the
1.2 Jy survey to galaxy fluctuations inferred on the 1/h Mpc scale in
the same survey by nearly a factor of 5!  On the other hand, CDM
models with $\Omega h \sim 0.2$ are remarkably successful at fitting
observations of large-scale structure\cite{peacock}.

In a flat universe, topological defect models also fail to produce the
observed large-scale structure.  Albrecht and Stebbins\cite{alb92}
found that the predicted spectrum of density fluctuations in a cosmic
string model is peaked on the string coherence scale, which is even
smaller than the horizon size.  Because of this lack of large-scale
power, they concluded that a string-seeded CDM model was not
compatible with the observed large-scale structure and failed more
dramatically than the inflationary scenarios.  In Pen, Spergel and
Turok\cite{pen94a}, hereafter PST, we explored global defects such as
global strings, monopoles, texture and non-topological textures.
While monopoles, texture and non-topological textures all predict a
power spectrum of density fluctuations with more large-scale power
than the cosmic string model, these models still fail to produce the
observed level of galaxy fluctuations by nearly an order of magnitude
on the 20/h Mpc scale.  This failure is due to the defect coherence
scale at equality being too small.  As this scale is proportional to
$\Omega h^2$, this problem will be alleviated in a low $\Omega$
universe.

In the excitement following the announcement\cite{smo92,wri92} that COBE
has detected thermal fluctuations in the cosmic microwave background
(CBR), many cosmologists declared that the detection was evidence for a
flat universe and the inflationary scenario.  This, however, need not be
the only interpretation of the COBE results.  The two-year COBE
results\cite{smo94} are also consistent with open universe models with
either adiabatic density fluctuations\cite{kam94,rat94} or
equation-of-state density fluctuations\cite{spe93}.

Inflationary scenarios predict either a scale-invariant spectrum of CBR
fluctuations or a spectrum of fluctuations that has more power on large
angular scales\cite{bon93}.  In models with significant gravity wave
contribution\cite{bon93} and in power law inflationary
models\cite{ada92}, the low multipole fluctuations are enhanced relative
to fluctuations on small angular scales.  Inflationary models with
cosmological constant\cite{kof85} also predict enhanced contributions on
large angular scales.  On the other hand, analysis of COBE's two year
data set\cite{ben94,smo94} suggest that the low multipoles are not
enhanced but suppressed relative to the fluctuations on smaller angular
scales.  While there is still significant statistical uncertainty in the
two-year COBE data, the Tenerife CBR observations\cite{ten} and the FIRS
results\cite{Ganga} also hint that the slope of CBR fluctuations may be
steeper than predicted in any inflationary models.  This steep CBR
fluctuations spectrum is one of the predictions of open universe
models\cite{spe93}.  If the suppression of the low multipoles is still
seen in the four-year data, then the COBE data may turn out to be
incompatible with most inflationary scenarios.

If the universe is open, then topological defect models are particularly
attractive.  The basic concept of inflation lies in solving the monopole
and horizon problems through an extended de-Sitter phase.  As other
authors\cite{got83} have pointed out, the curvature of space-like
section in a de-Sitter and also an empty universe depends on the choice
of coordinate systems, and could be either flat or hyperbolic.  The
prime reason for considering a flat universe a generic prediction of
inflation is the presence of the flatness problem.  In this paper we
will propose a solution along the lines of Linde, by using the weak
anthropic principle as a selection effect.  It hinges on the huge photon
to baryon ratio (approximately $10^{10}$), and gives a natural scale to
the problem.  The weakest point of inflation has been the fine tuning of
parameters required to generate the observed potential fluctuations.  So
even within the frame work of inflation, topological defects may still
be desirable as they explain this fine tuning more naturally.

But quite independently of the existence of inflation, if we are left to
find another route for explaining the large-scale homogeneity of the
universe, it would be most natural to assume that the universe started
with smooth initial conditions and that causal physics generated density
fluctuations.  Spergel\cite{spe93} presents analytical arguments that
suggested that defect models are likely to be compatible with
observations of large-scale structure in an open universe.  In this
paper, our numerical calculations support these arguments and suggest
that these models merit careful consideration.

This article builds upon the analytical and numerical techniques
presented in PST.  In PST, we attempted to quantitatively compare the
predictions to topological defect models in a flat universe to the
COBE observations and observations of large-scale structure.  This
paper extends our approach to an open universe.  Our CBR calculations
make no assumptions about the nature of the dark matter.  However,
when we compare the COBE normalized theories to the observed
large-scale structure, we assume that the universe is dominated by
cold dark matter.

In this paper, we estimate the amplitude of CBR fluctuations generated
by various topological defect scenarios.  Because of ease of
computation, we focus on the texture models.  While we have deferred
the challenge of evolving a string network in a hyperbolic universe,
we believe that qualitatively the results of this paper can be
extrapolated to other defect models.  In section~\ref{sec:open}, we
extend the analytical formalism that we developed in PST to open
universe models.  In appendix~\ref{app:vacuum}, we also extend the
formalism to vacuum-dominated models. In appendix~\ref{app:closed} we
extend to formalism to closed models.  In section~\ref{sec:numeric},
we describe our numerical algorithms for computing defect evolution.
In section~\ref{sec:results}, we present our numerical results and
emphasize the characteristic CBR signature of defects in open universe
models and then we discuss the predictions of the COBE-normalized open
universe models for large-scale structure.  In
section~\ref{sec:speculations}, we discuss philosophical motivations
for open universes.  In section~\ref{sec:conclusions}, we sum up.

\section{CBR Fluctuations in an Open universe}
\label{sec:open}

In this section, we modify the formalism developed in PST so
that we can calculate the amplitude of CBR fluctuations produced
in an open universe.

The assumptions entering in the calculations are the weak field limit
for gravity, which allows us to treat gravitational perturbations in
linearized form.  The quantum field is evolved as a classical field,
as it is a boson field with large occupation
number.  An additional assumption is the stiffness of the source term,
that gravity does not affect the evolution of the defects.  This is
certainly justified for global defects, which dissipate their energy
into Goldstone modes.  In the case of gauged cosmic strings one needs
to account for their energy dissipation in gravity waves.  The defect
field is governed by the nonlinear $\sigma$ model, which is a highly
nonlinear evolution equation.  While certain scaling laws can be
computed in a flat space time\cite{tur89}, this is not possible in the
transition regime between matter and curvature domination.  The field
correlation length can not be accurately specified, and most analytic
approaches are no longer applicable.  Thus one needs to simulate the
field evolution numerically, and measure these quantities.

The gravity is, however, still linear, and we will show below how to
solve that problem in an open universe once we are given the source
terms.  They are still straightforward integrals over the energy
momentum tensor with certain Green's functions.

In PST, we decomposed variations in the source stress energy tensor
and fluctuations in the metric into scalar, vector and tensor
components.  Variations in the trace of the spatial stress energy
tensor generates growing density modes that can form galaxies.  The
traceless scalar source term, the vector source term and the tensor
source terms do not generate growing density modes, however, they do
source decaying metric fluctuations that produce CBR fluctuations.  We
have reexamined our flat universe simulations and found that the CBR
fluctuations generated by the scalar modes are the dominant source of
fluctuations: the scalar growing mode term alone accounts for 70\% of
the CBR fluctuations.

The decomposition into scalar, vector and tensor modes is a non-local
calculation that is numerically challenging in an open universe.  In
order to evaluate the viability of defect models in an open universe,
we will focus on only the contributions of the growing scalar mode in
this calculation.  In an open universe, we expect that the vector and
tensor modes are even less important than in a flat universe as they
are suppressed relative to the scalar modes by powers of $(v/c)$,
where $v$ is the defect velocity.  In an open universe, the rapid
expansion of the universe slows the defect velocities.  Because of our
ignoring the anisotropic stress, vector and tensor modes, the
amplitude of CBR fluctuations calculated in this paper should be
multiplied by a factor between $1 - 1.2$.

In PST, we showed that the variations in the trace of
the spatial stress, $\Theta$, sources variations in
the scalar piece of the metric [PST 46]:
\begin{equation}
\ddot h^- + 2 {\dot a \over a} h^- = -8\pi G \Theta \equiv S
\label{spergel1}
\end{equation}
These metric fluctuations contribute to the Sachs-Wolfe integral:
\begin{equation}
\left({\delta T \over T}\right) = -{1 \over 2} \int_i^f d\eta
\dot K(x(\eta),\eta)
\label{spergel2}
\end{equation}
where
\begin{equation}
K = ({h^- \over 3} - {\cal J})
\label{spergel3}
\end{equation}
[PST 50] and
\begin{equation}
\dot {\cal J} + 2 {\dot a \over a} {\cal J} = -{1 \over 3} h^-
\label{spergel4}
\end{equation}
[PST 49].
Here, $a$ is expansion factor and dot denotes derivative with respect
to conformal time.  In the notation of PST, $J = k^2 {\cal J}$.

After a little algebra, equation (\ref{spergel1}), (\ref{spergel3}) and
(\ref{spergel4}) can be combined
to yield an equation for the scalar stress contribution to
CBR fluctuations:
\begin{eqnarray}
\dot K(x,\eta) = \int_0^\eta d\tilde \eta S(x,\tilde \eta) &&\left[
{a(\tilde \eta)^2 \over a(\eta)^2} - {\dot a(\eta) \over a(\eta) }
a(\tilde \eta)^2 E(\eta,\tilde\eta) \right.
\nonumber\\
&&- \left. \left({\ddot a(\eta) \over a(\eta)} -
3 {\dot a(\eta) \over a(\eta)^2}\right) {a(\tilde \eta)^2 \over a(\eta)^2}
H(\eta,\tilde \eta)\right]
\label{spergel5}
\end{eqnarray}
where
\begin{equation}
E(\eta,\tilde \eta) = \int_{\tilde \eta}^{\eta}
{d\bar\eta \over a(\bar \eta)^2}
\label{spergel6}
\end{equation}
and
\begin{equation}
H(\eta,\tilde \eta) = \int_{\tilde \eta}^\eta a(\bar \eta)^2 d\bar \eta
\int_{\tilde\eta}^{\bar \eta} {d\eta' \over a(\eta')^2}
\label{spergel7}
\end{equation}
Note that $E$ can also be used to relate the vector stress energy
source term to the vector metric fluctuations term [PST 47]
and to evolve the decaying scalar metric fluctuations
[PST 49].

In a flat matter-dominated universe, $a(\eta) = \eta^2$, thus,
$E(\eta,\tilde \eta) = 1/3\eta^3 - 1/3\tilde\eta^3$ and $H(\eta,\tilde\eta) =
\eta^5/15\tilde\eta^3 -\eta^2/6 + \tilde\eta^2/10$.  Combining
these results with equation (\ref{spergel5}) yields:
\begin{equation}
\dot K(x,\eta) = \int_0^\eta d\tilde \eta S(x,\tilde \eta)
\left({\tilde \eta^6 \over \eta^6}\right)
\label{spergel8}
\end{equation}
This simple result is due to the simple form of $a(\eta)$.

In an open matter-dominated
universe, the expansion factor has a more complicated form:
\begin{equation}
a(\eta) = {\Omega_0 \over -{\cal K}}
\left[\cosh(\sqrt{-{\cal K}}\eta) -1 \right]
\label{spergel9}
\end{equation}
where $\Omega_0$ is the density in matter today and ${\cal K}$ is the
curvature scale.  For the rest of the paper, we set ${\cal K} = -1$ and
use it as the physical length scale in the calculation.  Note that for
small $\eta$, equation (\ref{spergel9}) approaches the flat space form.
Combining equation (\ref{spergel9}) with equations (\ref{spergel6}) and
(\ref{spergel7}) yields:

\begin{equation}
E(\eta,\tilde \eta) = \left. {\dot a(\bar\eta) \over 3 a(\bar \eta)^2}
(1 - a(\bar \eta))\right|_{\bar\eta=\eta}^{\tilde\eta}
\label{spergel10a}
\end{equation}
and
\begin{equation}
H(\eta,\tilde \eta) = {1 \over \tilde a} - {5 \over 6} + {a^2 \over 6}
- {a \over 3} +
{\dot a(\tilde \eta) \left[1 - \tilde a  \right]\over 6 \tilde a^2}
\left[\dot a(\eta) (a-3) + 3(\eta - \tilde \eta)\right]
\label{spergel10b}
\end{equation}
where $\tilde a = a(\tilde \eta)$ and $a = a(\eta)$.
Equations (\ref{spergel5}), (\ref{spergel10a}) and (\ref{spergel10b})
can now be combined to yield
\begin{eqnarray}
\dot K(x,\eta) = \int_0^\eta d\tilde \eta S(x,\tilde \eta)
&&\left\{ {\tilde a^2 \over a^2}\left[1 + {2 -a -a^2 \over 3} + \left(
{5 \over a} + 2\right) \left({a^2 \over 6} - {a\over 3} - {5 \over 6}\right)
\right]  \right. \nonumber\\
&&+ \tilde a(1- \tilde a) \left[{-\dot a \over 3 a} + \left({5 \over a}+2
\right) {\dot a(a-3) +3\eta \over 6 a^2}\right] \nonumber\\
&&+ \left. \left[\tilde a - \dot {\tilde a}(1 - \tilde a) \tilde\eta\right]
{5 + 2 a \over a^3} \right\}
\label{spergel11}
\end{eqnarray}
In the limit of small $\eta$, equation
(\ref{spergel11}) reduces to equation (\ref{spergel8}).

The vector modes still allow a simple integral, for which we have
\begin{equation}
\dot{h}_i^V=\frac{16\pi G}{a^2}\int a(\eta')^2\Theta^V_id\eta'.
\label{eqn:vector}
\end{equation}
Tensor modes, on the other hand, propagate and require two Green's
functions to express.  As in [PST 54], we write
\begin{equation}
h^T_{ij} = 16\pi G \int \left[ \frac{G_1(\eta') G_2(\eta)
-G_2(\eta')G_1(\eta)}{W(\eta')}\Theta^T_{ij}(\eta)\right] d\eta'
\label{eqn:tensor}
\end{equation}
where now the Green's functions are
\begin{eqnarray}
G_1&=&\frac{\cos(k\eta)}{a}-\frac{\sin(k\eta)\dot{a}}{2a^2}
\nonumber\\
G_2 &=& \frac{\cos(k\eta)\dot{a}}{2a^2}+\frac{\sin(k\eta)}{a}
\nonumber \\
W&=&G_1\dot{G}_2-G_2\dot{G}_1
\label{eqn:ogreen}
\end{eqnarray}
We have assumed a decomposition in terms of eigenfunctions of the
Laplace-Beltrami operator $\nabla^2$, which in our framework will be
the sum of two sine frequencies and one Bessel function index.
Note that the tensor mode Green's function is equation
(\ref{eqn:ogreen}) are valid for open, flat and closed models.

In the next section, we present our numerical algorithms for evolving
the defect field to compute $S(x,\tilde\eta)$, of integrating equation
(\ref{spergel11}) to compute the metric fluctuations and for following photon
trajectories to integrate equation (\ref{spergel2}).

\section{Numerical Implementation}
\label{sec:numeric}

In this section we describe the numerical implementation issues.  The
original program is freely available by anonymous {\it ftp} from {\tt
astro.princeton.edu} in /upen/StiffSources/openuniverse.  It is
written in standard C++ and C, and should compile and execute on any
machine with these compilers.  It is optimized to execute very
efficiently on the convex vector architecture.  In fact, simulations
are always limited by memory, because the volume of a hyperbolic
universe is exponentially large, so the calculation scales as $O(N
\log(N))$ where $N$ is the memory requirement.

The basic strategy will be to apply the mode decomposition from PST.
In order to work in an open universe, many changes need to be applied
which are described below.

\subsection{Grid}

The very first obstacle is the formulation of a regular lattice to
discretize a hyperbolic manifold.  The requirements are: 1. it must have
constant volume per lattice element, 2. appear locally Euclidean,
3. be easily mapped onto the serial storage of a computer, 4. allow
the Laplacian to be easily invertible.

For this purpose the Poincar\'e metric provides a very nice tiling,
which retains many of the regularities of a flat space Cartesian
lattice.  The spatial metric is given by the line element

\begin{equation}
ds^2 = \frac{dx^2+dy^2+dw^2}{w^2}.
\end{equation}

With a change of variables $z=\ln(w)$, we satisfy all the requirements
stated above.  On small scales, it is explicitly Euclidean, which
simplifies the implementation of the differential operators.  This
metric maps onto the more familiar Friedman coordinates through the
change of variables

\begin{eqnarray}
x&=&\frac{\cos(\phi)\sinh(\chi)}{\cosh(\eta-\chi)} \nonumber\\
y&=&\frac{\sin(\phi)\sinh(\chi)}{\cosh(\eta-\chi)} \nonumber\\
z&=&\frac{\cosh(\eta)}{\cosh(\eta-\chi)} \nonumber\\
\cos(\theta)&=&\tanh(\eta)
\end{eqnarray}

The mapping onto the Friedman coordinates is shown in
figure~\ref{fig:pointof}.  The inverse mapping is displayed in
figure~\ref{fig:ftopoin}.  The salient features of this metric are the
explicit translational symmetry along all three dimensions, and the
rotational symmetry about the $z$ axis.  The only explicit numeric
anisotropy occurs for rotations in the $z-x$ and $z-y$ planes.  This
effect is easily tested for in the simulations by checking the
alignment of the quadrupole with the coordinate grid.

\subsection{Technical Issues}

For simplicity, we will work in units where $\Delta t=\Delta z=1$.
The horizontal discretizations $\exp(-z)\Delta x, \ \exp(-z)\Delta y$
are adjusted to be as close to unity as possible, while still
satisfying the periodicity constraints.  In these units our free
discretization parameters are the curvature radius $R$, the mesh
height $H$ and the periodicity length $L$ at $z=0$.  We choose
$-H/2<z<H/2$.  We divide the computational grids into tiers $T_i$ at
constant $z_i$, each of which is a square matrix.  We need to
subdivide $T_i$ into an integral number of lattice points, for which
we calculate the integer such that the area bounded by each lattice
point most closely approximates $dz^2$.  Since the $T_i$ are
represented by regular matrices, all parallelizations and
vectorizations are performed at this level.  The tier concept is then
implemented as a C++ class.  The key operation is the projection
(interpolation) of one tier into the geometry of its neighboring tier
above or below,

\begin{equation}
P_+: T_i \longrightarrow T_{i+1}, \ \ \ P_-: T_i \longrightarrow
T_{i-1} .
\end{equation}

In order for the discretization errors to be small, one needs to have
several grid cells per curvature length.  Another limiting constraint
is the scaling behavior.  In flat space, many defects achieve a
scaling solution where the energy density is proportional to $1/a^3$.
We require the numerical solution to achieve such a scaling law before
the horizon size grows to the curvature scale.

\subsection{Mode decomposition}
Our basic tool is the Fast Fourier Transform (FFT).  Since the grid
is periodic in $x$ and $y$, we can write any function $\phi$ as
a sum of Fourier components,
\begin{equation}
\phi(x,y,z)=\sum_{n,m} \exp\left[\frac{2i\pi (nx + my)}{L}
\right] \phi(n,m,z).
\end{equation}
In order to retain discrete orthogonality, the numerical grid points must
be all aligned at the phase origin $x=y=0$.  Then we simply keep a different
number of Fourier modes at each $z$.  The Laplacian then becomes a
second order ordinary differential equation in $z$,
\begin{equation}
\nabla^2\phi=-\frac{4\pi^2(n^2+m^2)}{L^2}\phi
+e^{-z}\frac{\partial}{\partial z}( e^z\frac{\partial\phi}{\partial z})
\end{equation}
which we integrate to second order accuracy.

In terms of the discretized variables,
\begin{eqnarray}
\nabla^2\phi&=&2\frac{\cos(\pi n/n_k)+\cos(\pi m/m_k)-2}{L}\phi
\nonumber \\
&&+\frac{\exp(-z_k)\left[\exp(z_{k+1/2})(\phi_{k+1}-\phi_{k})
-\exp(z_{k+1/2})(\phi_{k}-\phi_{k-1})\right]}{\Delta z^2}.
\label{dfd}
\end{eqnarray}

Note that the maximal mode $n,m$ depends on the level of the tier.  In
the case that the tier above does not contain a corresponding mode, we
use a zero boundary condition.  Equation (\ref{dfd}) is a tridiagonal
system which is solved in linear time.

Some care needs to be taken with the boundary conditions.  One can
easily violate causality from the non-local inversion of the Poisson
operator.  We thus need a boundary condition consistent with
causality.  A simple approach would be to set the boundary at the
edges of the computational domain, $z_u,z_l$ to zero.


In analogy with electro-magnetism, the zero boundary conditions
can be physically interpreted as a distribution of surface charges
which cancel the desired fields.  They cause waves to be reflected
at the boundaries, so that the gravitational field boundary conditions
become consistent with the field evolution.

The moving boundary condition can source scalar field and gravity
waves, but since these only travel at the speed of light, a
buffer zone will prevent them from affecting the photon cone
in the calculations.

\subsection{Field Evolution}
We implement a nonlinear $\sigma$ model following
the same approach as used in PST.  The equation of motion
for the continuum field reads:
\begin{equation}
\frac{1}{a^2}\partial_\eta a^2 \partial_\eta\phi = \nabla^2\phi+ \lambda \phi
\end{equation}
where $\lambda$ is a Lagrange multiplier which must be chosen to
satisfy the constraint $\phi^2=1$.  The time discretization has two
degrees of freedom, corresponding to the initial values of $\phi$ and
$\dot{\phi}$, which we represent through the field configuration
at two consecutive time steps.  We proceed in two steps.  First we
calculate the Laplacian.  Then we advance the field in the direction
of the Laplacian subject to the constraints of the nonlinear $\sigma$
model.  We treat the two issues in turn.

In the Poincar\'e metric, the Laplacian is expressed as
\begin{equation}
\nabla^2 \phi = \exp(-\frac{2z}{R}) (\phi_{,xx}+\phi_{,yy})
+\exp(-\frac{z}{R})\partial_z \exp(\frac{z}{R})\phi_{,z}.
\end{equation}
The first two terms are trivial to calculate with the standard central
difference formula.  With help of the projection operators $P_{+,-}$ we
can easily evaluate the vertical derivative
\begin{equation}
\exp(-\frac{z_k}{R})\left[\exp(\frac{z_{k+1/2}}{R}) (P_- \phi_{k+1} - \phi_k)
-\exp(\frac{z_{k-1/2}}{R}) (\phi_k-P_+ \phi_{k-1})\right]
\end{equation}
Along the $x,y$ axes we can simply use periodic boundary conditions.
At the top of the grid $z=H/2$ we simply extend our grid upward, which
costs very little in computational effort or memory because very
little volume is enclosed in that region.  The bottom boundary needs
to be treated more carefully.  We choose a safety buffer zone of a few
grid cells below the last grid point that is traversed by photons.



\subsection{Implementation}

For texture models, there is reason to believe that the primary
contribution comes from the spatial trace of the energy momentum
tensor,
\begin{equation}
\Theta=\sum_{i=1}^3 T^i_i
\label{eqn:trace}
\end{equation}

In flat and empty space, the exact texture solution generates an
energy momentum tensor which is such a pure trace.  As we describe
below, the problem is isomorphic to such a flat problem in both the
early and late time limits.  The main contribution thus arises from
the trace.  In our current implementation, we have chosen the
approximation to only retain the trace part.  For other defects,
especially the cosmic strings, other components of the energy momentum
tensor are expected to play a dominant role and one needs to implement
the full mode decomposition described above.

The simulations are constrained on several ends.  We need a fair
number of grid cells per curvature radius in order to achieve an
accurate flat scaling density for the field before the numerical
horizon size becomes comparable with the curvature scale.  A violation
of this constraint would cause the simulation to enter the curvature
transition with an incorrect energy density, which would appear as a
systematic subsequent error.  In practice this corresponds to about
8 grid cells per curvature radius.

With the 2 Gigabytes of memory on our convex C3440 we can run
simulations down to $\Omega=0.2$, which take about 4 hours of (wall
clock) execution time.

\subsection{Tests of the Code}
The two extreme limits of the parameter space have exact solutions.
When $\Omega=1$, we have an expanding flat space and we can test
the exact scaling solution for a single unwinding texture as we
did in the flat space calculations\cite{pen94a}.  The other limit
$\Omega=0$ is an empty universe, which is nothing more then Minkowski
space.  Using the Milne transformation, we map the exact scaling
solution for a texture as initial condition on our grid, and test
for the subsequent evolution.  In an empty universe using conformal
coordinates, the
cosmic scale factor $a=\eta$, so the Milne solution to the Einstein
equation in an empty universe becomes
\begin{equation}
ds^2=-d\tau^2+\tau^2(\frac{dq^2}{1+q^2}+q^2d\Omega^2)
\end{equation}
With a change of coordinates $r=q\tau, \ \ t=\tau\sqrt{1+q^2}$ we recover
the Minkowski metric $ds^2=-dt^2+dr^2+r^2d\Omega^2$.
Our numerical grid performs very well on this test.  Since the
same code has very small errors on both limits, we can claim some
confidence that it should perform well in between.

\section{Results}
\label{sec:results}

\subsection{CBR fluctuations}

Using the algorithm outlined in the previous section, we have computed
the CBR fluctuations produced by scalar potential fluctuations in an
open universe.  The results of these calculations are shown in figures
\ref{fig:omega21}, \ref{fig:omega4} and \ref{fig:oflat} for four
different observers in $\Omega = 0.21, 0.4,$ and $\Omega = 1$
universes.  The anisotropy of the grid is almost visible as an
enhancement of the quadrupole in figure \ref{fig:oflat} for
$\Omega=1$.  In this case the light rays always move at a constant
angle to the grid, so we conclude that even in the worst case, the
grid anisotropy only has a minor effect.

In an open universe, the defects dynamics slows down due to the rapid
expansion of the universe, which exhibits itself as a loss of power on
large angular scales, in particular the quadrupole terms.

The different lines in the figure denote the results from different
realizations and the spread in values is a measure of the variance in
$c_l$.  The abscissa in the plot is the amplitude of the multipole
moments, $c_l = \sum_m a_{lm}^2/(2l + 1)$, weighted by $l(l+1)$.  In a
flat inflationary model with scale-invariant spectrum of fluctuations,
$c_l l (l+1)$ is a constant for $l << 200$.  Our results imply that for
defect models in a low $\Omega$ universe, the shape of the multipole
spectrum is qualitatively different from other models.  The low order
multipoles are strongly suppressed, while the multipoles on scales
significantly below the angular scale subtended by the curvature scale
today still have a scale invariant form.

For purposes of comparison with the DMR measurements of the temperature
fluctuations, we have fit the results of our numerical simulations with
a fitting form:
$$c_l l (l+1) = c_0/{\left[1+(l_{max}/l)^{2.5}\right]}^{0.4}$$
The simulations are fit by $l_{max} \simeq 3 \Omega^{0.5}$.  Even if the
universe is flat, the quadrupole is somewhat suppressed in any model
with topological defects as defects on scales comparable to and greater
than the horizon size have not yet had time to collapse.


The qualitative features of topological defects in an open universe can
be approximated using any flat universe calculation.  To first
approximation, one can consider the photon sphere we observe today
projected back at a redshift of $1+z \approx 1/\Omega$.  This will also
produce a flat spectrum for large $l$, with a white noise cutoff for
small $l$.  We used our flat space stopped at $z=1$ and $z=1.5$, and
the multipole spectrum is depicted in figures \ref{fig:flatz1} and
\ref{fig:flatz15}.

The COBE two year observations have only been analyzed under the
assumption of a power law spectrum of CBR
fluctuations\cite{gorski,ben94,smo94}.  These analyses conclude that
COBE has measured a quadrupole of only 6 $\pm 3 \mu$K while their fit to
larger multipoles imply $Q_{rms} = 20 \mu$K.  In an inflationary model,
such a small value for the quadrupole should be observed in less than 5\%
of the universe.  In a defect model in an open universe, the low value
for the quadrupole is predicted.  On the other hand, COBE two-year
measurements do not find a suppression of the $l=3$ and $l=4$ modes.

As discussed in section 3, our calculations only include the contributions
from the scalar growing mode to the CBR fluctuations.  The calculations do
not include contributions from decaying modes, vector fluctuations and
gravity wave fluctuations.  These contributions are subdominant in a flat
universe and should be even smaller in an open universe.  If these
terms were included, then the amplitude of the CBR fluctuations
would be increased by a factor $f \sim 1.0 - 1.4$.  The upper bound
is based on our flat space calculations\cite{pen94a}.  Thus, the amplitude
of CBR fluctuations needs to be multiplied by this factor.

Topological defect theories have one free parameter: the scale of
symmetry breaking, $\phi_0$.  Note that the abscissas in figures
\ref{fig:omega21} - \ref{fig:oflat}
need to be multiplied by $8\pi^2 G\phi_0^2$ to convert the results of
the calculations into temperature fluctuations.  We will fix this
parameter by normalizing our results to the COBE two-year observations
by convolving our results with a Gaussian beam with full width half
maximum of 10$^o$ and fixing $(\delta T/T)^2_{rms}$ to the value of 40
$\mu$K suggested by harmonic analysis of the two year data\cite{gorski}.
The second column in table 1 shows the normalized value of $8\pi^2
G\phi_0^2$ for different values of $\Omega$.  Note that these values are
slightly higher than our earlier calculation based on the DMR one-year
data.

While our calculation did not compute the amplitude of CBR fluctuations
on small angular size, we can extrapolate calculation on defects in a
flat universe to the open case.  Coulson et al\cite{coulson94} found
that in a re-ionized universe, the multipole spectrum was flat from large
angular scales to $l\sim 60$.  This multipole moment corresponds to the
angular size subtended by a texture collapsing near the surface of last
scatter in a re-ionized universe.  In an open universe, the relationship
between horizon size and angle is altered: $\theta \sim \Omega^{0.5}
z^{-0.5}$ for $z >> \Omega^{-1}$.  Thus, we expect that in an open
re-ionized universe, the multipole spectrum would be flat from $l \sim
l_{max}$ to $l \sim 60 \Omega^{-0.5}$.

If the early universe was not re-ionized by a generation of star
formation before $z\sim 50$, then CBR observations on small angular
sizes are probing the universe at $z\sim 1300$, the epoch of
recombination.  In both defect models and scale-invariant curvature
models, there should be a ``Doppler peak'' at $l\sim 200 \Omega^{-0.5}$.
In an inflationary model with curvature fluctuations, this peak is
produced by the sum of velocity perturbations and the product of
potential fluctuations with entropy fluctuations.  In these models, the
fluctuations are all produced by the growing modes.  In a defect model,
the ``Doppler peak'' is due primarily to the entropy and potential
fluctuations produced by the decaying modes excited by the collapse of
defects \cite{spergel95}.  As in curvature models, this peak should
occur at $l \sim 200 \Omega^{-0.5}$, thus measurements of its location
may provide a determination of $\Omega$.

Just as in a flat universe, one of the distinctive predictions of a
defect model are the non-Gaussian character of the temperature
fluctuations.  This non-Gaussianeity should be most apparent not in the
distribution of temperature fluctuations, but in the distributions of
temperature gradients \cite{coulson94,branden93}.  In a flat universe,
Coulson et al.  find that this non-Gaussian character is most apparent
on angular size of $\sim 3^o$, the angular size subtended by the surface
of last scatter in a re-ionized universe.  In an open universe, this
angular size is shifted to $\sim 3 \Omega^{0.5}$ degrees.

While simulations were done only for texture models, the results can easily
be extrapolated to other defect models in an open universe.  In all of these
models, we expect a similar suppression of the low multipoles as the rapid
expansion of the universe slows the evolution of the defects responsible
for generating fluctuations on large angular scales.

\subsection{Large Scale Structure}

Having normalized the defect model to fluctuations in the CBR, we now
turn to predictions for mass fluctuations on the scale probed by
galaxy surveys.  Fluctuations on scales smaller than $\sim 1000$ Mpc
were generated when $\Omega$ was close to unity, thus, the results
of our earlier work on density fluctuations in a flat universe can
be directly extrapolated to open models.

In Pen et al.\cite {pen94a},  density fluctuation
computed by numerical simulations
fit by a function [PST 32, 33]:
$$P(k) = {D(\Omega)^2  \over \Omega h^2} {\alpha k \over
[1 + (\beta k) + (\gamma k)^{1.5}]^2}$$
where $D(\Omega) \simeq \Omega^{0.7}$ is ratio of the linear growth
to that in an $\Omega = 1$ universe, $\alpha = 225 (\epsilon/3.7
\times 10^{-4})^2/(\Omega h^2)$,
$\beta = 3.5/(\Omega h^2)$ Mpc, and $\gamma = 2.75/(\Omega h^2)$ Mpc.
Using the normalization in Table 1, this directly yields
the amplitude of mass fluctuations predicted by linear theory.

On the scale of tens of kiloparsecs, light clearly does not trace mass.
This is the source of the missing mass problem.  It is less certain
whether light traces mass on the scale of several megaparsecs.  Cosmologists
parameterize this uncertainty by a bias parameter, $b$, the ratio
of the variance in the fluctuation in the galaxy counts to the fluctuations
in the cosmic density field.  Here, we determine the value of $b$
needed to fit astronomical observations.

The QDOT survey measured the fluctuations in galaxy counts by obtaining
redshifts to a large infrared selected galaxy sample.  Saunders et
al.\cite{saunders} smoothed their galaxy density field with a Gaussian
smoothing window with filter length of 5$h^{-1}$, 10$h^{-1}$, and
20$h^{-1}$ Mpc and found a variance of $0.436 \pm 0.091$, $0.184 \pm
0.05$ and $0.0669 \pm 0.019$ in the density field.  The fourth, fifth
and sixth column in Table 1 lists the required bias factor needed to fit
the central value in the QDOT survey.  The statistical uncertainties in
the QDOT survey and the COBE measurements lead to a $\sim 25\%$
$1\sigma$ uncertainty in $b$.  This is in addition to the uncertainty
due to the limitations of our numerical calculations.

In figure \ref{fig:apm}, we compare the predicted power spectrum of
density fluctuations to the power spectrum of galaxy fluctuations
inferred from APM survey.  In this figure, we assumed a bias
factor of $1.5$.

\section{Speculations}
\label{sec:speculations}

At various points in history, different choices for the curvature of the
universe have been considered most natural.  Einstein initially
considered an eternal and flat universe with a cosmological constant
most aesthetically pleasing.  But with Hubble's discovery of the
universal expansion, the common belief was that the universe should be a
closed three sphere, which is bounded in both space and time.  This
universe would end in a few Hubble times, and again we would live in a
very ordinary epoch.  We will use the same Copernican principle to argue
that an open universe is almost as well suited.

In the last decade it has become fashionable to return to consider
spatially flat universes at most appealing, since there would not be any
curvature scale which needs to be explained.  This is primarily due to
Dicke's anthropic argument, who used the Copernican principle to argue
that we should not be living just at he end of the flat epoch.  Not to
forget that the inflationary paradigm, which appeals to the de-Sitter
model to solve the so-called horizon problem, is simplest to explain in
a scale-free scenario of a flat universe.  But the very problem that
they are invented to solve, the absence of a preferred curvature scale,
leads to a clash with the Copernican principle as we live at a very
special time, just near the beginning of a matter-dominated universe
which would now last for a truly lengthy period of proper time.
As we have seen, in the absence of a perfect fluid to label a preferred
coordinate frame, the curvature of a universe can be transformed from
flat to hyperbolic by a gauge choice.  The same holds true for a
de-Sitter space.

An open universe model does have aesthetic advantages that have been at
times overlooked.  In an open universe, we are most likely to live in
the brief period of time between radiation and curvature domination.
Density fluctuations do not collapse during the radiation dominated
epoch and are growing logarithmically slowly during curvature
domination.  If a universe went directly from a radiation dominated
phase into curvature domination, no structure or life would ever be
conceivable.  It is only due to a lucky coincidence that we have a
slight baryon asymmetry of $\eta=10^{-11}$, possibly due to
baryogenesis in the electro-weak phase transition.  This allowed
structures to form through gravitational instability in the short
interval between matter-radiation equality and curvature domination.
Since this interval lasts only for a finite time, and we live
approximately in the center of this period, there is no violation of the
Copernican principle.  We thus appeal to the observed smallness of
electro-weak baryogenesis to set the scale for the hyperbolic curvature.
Dirac's small number is not really a single small number.  The radius of
curvature and the horizon size today are simply the product of the
proton mass, the baryogenesis photon to baryon ratio, and the smallness
of initial fluctuations, observed by COBE to be $10^{-5}$, which in the
topological defect framework arises from the GUT scale.  Such a model
has no fine tuned parameters.

{}From a geometric viewpoint, a closed universe is appealing due to its
simplicity: a three sphere is the unique universal covering of
positively curved three-manifolds, and has a finite volume which would
certainly be an attractive property for any designer or process which
might have created the universe.  But it is interesting to note that
there are only a finite number of alternate global topologies which
such a designer has to choose from.  The projective 3-sphere $P^3$ is
one such example.  If one analyzes negatively curved spaces, one can
of course consider the global covering $H^3$, but there are many
alternatives, including the periodic Poincare space which we utilized
in this paper.  The name ``open'' only applies to the local
properties, and we can certainly have a spatially closed ``open''
universe.  If one considers only hyperbolic spaces of finite volume,
one finds an infinite number of possible topologies.  At fixed
curvature $R_c$, the volumes of these topologies can have collection
points on the real line, and one might expect our universe to be
chosen from a topology near such a collection point.  A number of
authors have attempted to calculate transition probabilities between
these configurations in 2+1 dimensions\cite{wit85}.  These studies
suggest that topological change may well be possible.  Unfortunately,
as with most quantum  gravity calculations, many infinite quantities
arise in the process, making is difficult to uniquely predict the
outcome of such an estimate.

Whether the negative curvature results from quantum gravitational
tunnelling, or an alternate exit from the de Sitter phase, or some other
yet unknown means, the anthropic principle does set a minimum scale to
the curvature radius.  In order to bound it from above, one could argue
that the intrinsic process forms hyperbolic space-times with small
curvature, most of which are not observable.  So we might live in the
smallest allowed scenarios which allow nonlinear structures to form.

However, as we lack a theory of quantum gravity that can predict whether
a flat, open or closed universe is most likely, we believe that all of
these models merit careful consideration.  Ultimately, this question
must be resolved observationally.  We have shown that the philosophical
arguments that are been invoked to argue for a flat universe are quite
ambiguous, and could just as well be used to argue for an open scenario.

\section{Conclusions}
\label{sec:conclusions}

In this paper, we explored the cosmic microwave background signature of
the formation of large scale structure by defects in an open universe.
We have described a new efficient exact solution of the linearized
Einstein equations in hyperbolic FRW spacetimes.  The otherwise
expensive mode decomposition can be implemented very efficiently thanks
to the Fast Fourier Transform in the Poincar\'e metric.  This
formulation was then applied to calculate predictions of texture
models in an open universe.  We then addressed the classical
philosophical arguments including the flatness problem.  We showed that
the same anthropic and Copernican arguments that were used to argue for
a flat universe, are in fact better satisfied in an open model.  The
curvature scale is naturally explained as a product of three moderately
big numbers, which are experimentally well established.

For $\Omega = 0.4$ and $h = 0.7$, the power spectrum of density
fluctuations in a COBE-normalized texture model has the correct spectral
shape and is consistent with the observed level of galaxy fluctuations
for $b = 2 -4$.  The uncertainty in normalization is a combination of
the numerical uncertainties in our calculations and the statistical
uncertainties in the observations.  A model with $\Omega = 0.4$ and $b =
2$ is consistent with various dynamical measurements of $\Omega$ on the
scale of clusters and superclusters.  The observations that are most
problematic for this model are the large-scale streaming velocities
inferred from various proper distance surveys.

While our work focused on textures in an open universe, we expect
qualitatively similar results for other defect models.  The basic
results appear to be governed by geometry and the changed relationship
between angle and physical scale.  This is apparent when we compare a
flat universe simulation stopped at $1+z = \Omega^{-1}$ and rescaled by
a factor of $\Omega$ in angle with an open universe simulation.

Defects in an open universe make a distinctive prediction for the CBR
spectrum.  In these models, very few fluctuations are generated at late
times and at large angular scales.  Thus, the models predict a
suppression of the quadrupole and other low multipole moments.  The low
value of the quadrupole detected by COBE is consistent with this set of
theories.  A detailed analysis of the COBE DMR results is needed to
determine which range of values of $\Omega$ are compatible with the
observed universe.  A qualitative understanding is quite straightforward
and is obtained by rescaling the flat space spectra and introducing a
break near the curvature scale.

Defects in an open universe are a viable alternative to popular
scenarios for structure formation and merit closer study.

\acknowledgements

We would like to thank Neil Turok for helpful comments.  UP and DNS are
partially supported by NSF Contract Nos. AST 88-58145 and ASC 93-18185
(GC3 HPCC collaboration) and by NASA Contract No. NAGW-2448.  Numerical
calculations were performed on the Convex C3440, which was partially
funded by NSF Contract No. AST 90-20863.

\appendix{Geometric Formulae in Hyperbolic Spaces}

\section{Notation}

$R_c$ will denote the curvature radius of the universe, which is given
by $R_c=c/(H_0 \sqrt{1-\Omega})$.  A subscript of $0$ denotes a
parameter's present value.  We use $t$ to denote the proper time,
$\eta$ to measure conformal time.  They are related through
\begin{equation}
t=\frac{t_0}{\sinh\eta_0}(\sinh \eta - \eta)
\end{equation}
In conformally hyperbolic coordinates, the expansion factor is
$a=\cosh \eta - 1 = (2/\Omega) -2$.  The conformal time measures the
number of comoving curvature radii traversed by a photon.  It is given
by $\eta=\cosh^{-1}(2/\Omega-1)$.
We write the FRW metric as
\begin{equation}
ds^2=-dt^2+a(t)^2(\frac{dr^2}{1+r^2} + r^2 d\Omega^2).
\end{equation}
Proper distance between two time synchronized observers is
$\Delta s=a R_c \sinh^{-1} r$.  The coordinate $r$ is in units of
comoving curvature radii.

\section{Vacuum Dominated Universes}
\label{app:vacuum}
This appendix contains results for a vacuum-dominated flat universe
containing dust and vacuum energy.  In this universe,  the expansion
factor can be computed from the energy equation:
$$\left({da \over dt}\right)^2 = H_0^2 \left[\Omega_0 {a_0 \over a}
+ (1 - \Omega_0) {a^2 \over a_0}\right]
$$
where $\Omega_0$, $a_0$ and $H_0$ are the density in matter, the expansion
factor and the Hubble constant today and $t$ denotes physical time.
This equation can be solved to yield:
$$a(t) = {3 \over 2} \Omega_0^{1/3} (1 - \Omega_0)^{1/6} \sinh^{2/3}(H_0 t)
$$

The conformal time, $\eta$ in a flat vacuum dominated universe can
be computed with the aid of Gradshteyn and Rhyzik equation (3.166.22):
\begin{eqnarray}
\eta &=& \int_0^t {dt \over a(t)} = \int_0^a {da \over a(t) (da/dt)}
= \int_0^a {dx \over \sqrt{x(1 + x^3)}} \nonumber\\
&=& {1 \over 3^{1/4}} {\cal F}\left[
\cos^{-1}\left({1 + (1 - \sqrt{3}) a \over
1 + (1 + \sqrt{3}) a}\right), {\sqrt{2+ \sqrt{3}} \over 2} \right]
\label{dnsA3}
\end{eqnarray}
Here, ${\cal F}$ is the elliptic integral of the first kind.

It would appear that this complicated relationship between $a$ and
$\eta$ would make it impossible to evaluate equations (\ref{spergel6})
and (\ref{spergel7})
analytically.  However, by change of variables from $\eta$
to $a$, these equations
become remarkably tractable:
\begin{equation}
E(a,\tilde a) = \int_{\tilde \eta}^\eta {d\bar\eta \over a(\bar\eta)^2}
=\int_{\tilde a}^a {d\bar{a} \over {\bar a}^2 \sqrt{\bar a + \bar a^4}}
={2 \over 3} \sqrt{1 + a^{-3}}|_a^{\tilde a}
\label{dnsA4}
\end{equation}
and
\begin{eqnarray}
H(a,\tilde a) &=& \int_{\tilde a}^a {d\bar a \bar a^2 \over
\sqrt{\bar a + \bar a^4}} E(\bar a, \tilde a) \nonumber\\
&= &{2 \over 3} (\bar a - a) + {4 \over 15}\sqrt{1 + \bar a^{-3}}
\left[\bar a^{5/6} _2F_1({1\over 2}, {5\over 6};{11 \over 6}, -\bar a)
\right]_{\tilde a}^a
\label{dnsA5}
\end{eqnarray}
Here, $_2F_1$ is Gauss' hypergeometric function.  Gradshteyn
and Rhyzik equation (3.194.1) was used to compute (\ref{dnsA5}).  Equations
(\ref{dnsA4}) and (\ref{dnsA5}) can be used to compute CBR fluctuations in
a vacuum dominated model.  Note that (\ref{dnsA4})  and (\ref{dnsA5})
can also be used
to evolve vector and decaying scalar modes:
$$h_i^V(\eta) = 16 \pi G \int \Theta_i^V(\tilde \eta)  E(a,\tilde a)
\tilde a^2 d\tilde \eta$$

\section{Closed Models}
\label{app:closed}

In a closed matter-dominated FRW model, the scale factor is
\begin{equation}
a(\eta)=1-\cos(\eta)
\end{equation}
Thus, the scalar modes are given by equation (\ref{spergel5}) with
\begin{eqnarray}
E(\eta,\tilde{\eta}) &=& \left. \frac{\dot{a}}{3a}(1+a)
\right|_{\eta}^{\tilde\eta}
\nonumber \\
H(\eta,\tilde\eta) &=& \left.
(\frac{3\bar\eta}{2}-2\sin(\bar\eta)+\frac{\cos(\bar\eta)
\sin(\bar\eta)}{2})\right|_{\tilde\eta}^\eta \times E(\eta,\tilde\eta)
\end{eqnarray}
The vector modes still satisfy equation (\ref{eqn:vector}) and with
the appropriate form for $a(\eta)$, equation (\ref{eqn:tensor})
describes the evolution of tensor modes.

\newpage
\begin{figure}{Table 1}
Comparison with QDOT Observations

\begin{tabular}{|ccccc|}
\hline
 $\ \ \Omega_0$\ \ \  & $H_0$ && {Required Bias }  \hfil& \\
&km/s/Mpc & \  $5h^{-1}$Mpc & \ $10h^{-1}$Mpc & \ $20h^{-1}$ Mpc \\
\hline
  0.2 &   0.5& 3.8- 4.7& 4.4- 5.8& 5.3- 7.1 \\
  0.2 &   0.6& 3.1- 3.8& 3.6- 4.8& 4.5- 6.1 \\
  0.2 &   0.7& 2.6- 3.2& 3.1- 4.1& 4.0- 5.3 \\
  0.2 &   0.8& 2.2- 2.8& 2.7- 3.6& 3.6- 4.8 \\
  0.2 &   0.9& 2.0- 2.4& 2.4- 3.2& 3.3- 4.4 \\
  0.4 &   0.5& 2.8- 3.4& 3.5- 4.6& 4.8- 6.5 \\
  0.4 &   0.6& 2.3- 2.8& 3.0- 4.0& 4.3- 5.7 \\
  0.4 &   0.7& 2.0- 2.4& 2.6- 3.5& 3.9- 5.2 \\
  0.4 &   0.8& 1.7- 2.1& 2.4- 3.1& 3.6- 4.8 \\
  0.4 &   0.9& 1.5- 1.9& 2.2- 2.9& 3.4- 4.5 \\
  1.0 &   0.5& 2.3- 2.8& 3.4- 4.5& 5.7- 7.6 \\
  1.0 &   0.6& 2.0- 2.4& 3.1- 4.0& 5.3- 7.1 \\
  1.0 &   0.7& 1.7- 2.2& 2.8- 3.7& 5.0- 6.7 \\
  1.0 &   0.8& 1.6- 2.0& 2.6- 3.5& 4.8- 6.5 \\
  1.0 &   0.9& 1.5- 1.8& 2.5- 3.3& 4.7- 6.3 \\
\hline
\end{tabular}
\end{figure}

\newpage

\begin{figure}{Poincare to Friedman mapping}
\vskip 6pt
\epsfxsize=\hsize
\caption{
The solid lines are the equispaced surfaces of constant $z$.  The
dotted lines are geodesics of constant $x$.  The periodicity boundary
is selected along one of these curves.  The crosses indicates
the location of our numerical grid points, which are regular and
evenly spaced in the Poincar\'e metric.}
\label{fig:pointof}
\end{figure}

\newpage

\begin{figure}{Photon Trajectories}
\vskip 6pt

\epsfxsize=\hsize
\caption{This figure depicts various photon geodesics in the Poincar\'e frame.
The axes are in units of constant distance, so $\rho=\exp(-z)(x^2+y^2)$
and $z=ln(w)$.  The heavy line is the fudicial geodesic along which our grid
is periodic.  The radial lines is the trajectory traversed by photons.
The concentric lines are circles of constant distance from the origin,
i.e. spheres in the Poincar\'e frame.}
\label{fig:ftopoin}
\end{figure}

\newpage

\begin{figure}{$\Omega=0.21$ Multipole Spectrum}
\vskip 6pt
\epsfxsize=\hsize
\caption{The eight jagged lines correspond to observers in
different universes or at different locations.  Flat inflationary models
predict $c_l l(l+1)$ as constant.  The dashed line is a parametrization
of the CBR open universe spectrum:
$c_ll(l+1)\propto/(1+(l_{\rm max}/l)^q)^{1/q}$
with $l_{\rm max}=8$ and $q=2.5$.  In general $l_{\rm max}$ scales as
$\propto \Omega^{-1/2}$.}
\label{fig:omega21}
\end{figure}

\newpage

\begin{figure}{$\Omega=0.4$ Multipole Spectrum}
\vskip 6pt

\epsfxsize=\hsize
\caption{Same as figure \protect\ref{fig:omega21}, but for $\Omega=0.4$.}
\label{fig:omega4}
\end{figure}

\newpage

\begin{figure}{$\Omega=1$ Multipole Spectrum}
\vskip 6pt

\epsfxsize=\hsize
\caption{Nearly flat universe calculated using the open universe code.
Here we recover the flat Harrison Zeldovich spectrum.}
\label{fig:oflat}
\end{figure}

\newpage

\begin{figure}{Flat Space Approximation}
\vskip 6pt

\epsfxsize=\hsize
\caption{Flat space simulation at z=1}
\label{fig:flatz1}
\end{figure}
\begin{figure}
\vskip 6pt

\epsfxsize=\hsize
\caption{Flat space simulation at z=1.5}
\label{fig:flatz15}
\end{figure}

\newpage

\begin{figure}{Comparison with APM survey}
\vskip 6pt

\epsfxsize=\hsize
\caption{Comparison of the predicted open universe power spectrum to
the APM survey for various parameters of Hubble's constant and $\Omega$}
\label{fig:apm}
\end{figure}

\end{document}